\documentclass[useAMS,usenatbib,a4paper]{mn2e}
\usepackage{epsfig,bm}
\usepackage{amsbsy,amssymb,amsmath}
\usepackage{float}                    
\usepackage{subfigure}
\usepackage{natbib}
\usepackage[normalem]{ulem}

        \usepackage{graphicx}
        \usepackage{epstopdf}
        \usepackage{hyperref}

\defcitealias{Planck_blue}{Planck Collaboration 2005}
\defcitealias{Planck_cosmo}{Planck Collaboration 2013}

\title{Testing CMB polarization data using position angles}
\author[Michael Preece and Richard A. Battye]
{Michael Preece and Richard A. Battye\\
Jodrell Bank Centre for Astrophysics, School of Physics and Astronomy, The University of Manchester, Manchester, M13 9PL, U.K.}

\begin{document}

\maketitle

 \begin{abstract}

We consider a novel null test for contamination which can be applied to CMB polarization data that involves analysis of the statistics of the polarization position angles. Specifically, we will concentrate on using histograms of the measured position angles to illustrate the idea. Such a test has been used to identify systematics in the NVSS point source catalogue with an amplitude well below the noise level. We explore the statistical properties of polarization angles in CMB maps. If the polarization angle is not correlated between pixels, then the errors follow a simple $\sqrt{N_{\rm{pix}}}$ law. However this is typically not the case for CMB maps since these have correlations which result in an increase in the variance since the effective number of independent pixels is reduced. Then we illustrate how certain classes of systematic errors can result in very obvious patterns in these histograms, and thus that these errors could possibly be identified using this method.  We discuss how this idea might be applied in a realistic context, and make a preliminary analysis of the WMAP7 data, finding evidence of a systematic error in the Q and W band data, consistent with a constant offset in $Q$ and $U$.

\end{abstract}

\section{Introduction}
\label{Intro}

Measurements of the angular power spectrum of temperature anisotropies in the Cosmic Microwave Background (CMB) have had a significant impact on our understanding of the Universe (\citealt{DASI_detect}, \citealt{Maxima}, \citealt{MSAM}, \citealt{CBI}, \citealt{BOOMERanG}, \citealt{JB-IAC}, \citealt{CAT}, \citealt{ACBAR}, \citealt{VSA}, \citealt{SPT}, \citealt{ACT}). The recent WMAP and Planck data has constrained the 6 parameter $\Lambda$CDM model to high precision (\citealt{nine_maps}, \citealt{nine_cosmo}, \citetalias{Planck_cosmo}). The present frontier of CMB research is the measurement of polarization (\citealt{CBI_pol}, \citealt{DASI_pol}, \citealt{BOOMERanG_pol}, \citealt{third_pol}, \citealt{MAXIPOL}, \citealt{CAPMAP}, \citealt{QUaD}, \citealt{BICEP}). Not only is this significantly weaker, but the polarization field is spin-2 leading to the dichotomy between E and B modes (\citealt{gravwaves}, \citealt{CMBstat}). E-modes are created by all inhomogeneities, but B-modes can only be due to vorticity and gravitational waves at linearised order. Since primordial vorticity is not generated during inflation, detection of B-modes could be seen as an indirect detection of primordial gravitational waves, something which would constrain the energy-scale at which inflation took place. 

Measurement of the B-mode power spectrum will be technically challenging since a scalar-to-tensor ratio of $r~\approx~0.01$ will require measurements with r.m.s. noise levels below $100{\rm nK}$. In addition, contamination has to be constrained to be well below the noise level. This will come from three primary sources: astrophysical foregrounds (emission from diffuse components of the galaxy and extragalactic sources), the atmosphere and the telescope/receiver system. A variety of null tests have been applied to CMB measurements in order to refine and test the quality and consistency of the data. In this paper we put forward a test using polarization position angles.

The kind of tests we will be discussing are those which one might apply at the stage where one has a map or, in the case of an interferometer, visibilities. We will concentrate on the map based case in this paper, but it should be easy to apply similar ideas to interferometer visibilities. Tests of the Gaussianity have been applied to temperature anisotropies and have proved useful in the removal of a range of systematics, for example in the COBE data \citep{COBE_gauss}. Additional checks can be made in the case of polarization, in particular one can check that the cross-correlation power spectra $C_{\ell}^{EB}$ and $C_{\ell}^{TB}$ are consistent with zero. At present all published data is consistent with this standard hypothesis, which is a result of the parity of the B mode signal.

Our test will use the polarization position angle, $\alpha$, which can be defined in terms of the Stokes parameters $Q$ and $U$ by 
\begin{equation}
\alpha=\frac{1}{2}\tan^{-1}\left(\frac{U}{Q}\right)\, ,
\end{equation}
which can take values between $-90^{\circ}$ and $90^{\circ}$. Since this is a coordinate dependent quantity, it is not often discussed in the context of cosmology. However, it is commonly used in the context of other astrophysical sources and, in fact, our motivation for this test comes from previous work which has used similar techniques to establish the existence of biases (at the level of $\langle\Delta\alpha^2\rangle^{1/2}\approx 0.3^{\circ})$ in the polarization position angles measured in the NRAO-VLA Sky Survey (NVSS) point source catalogue \citep{NVSS_bias}. Briefly, it was found that there were highly significant biases in the measured position angles toward angles that are integer multiples of $45^{\circ}$ (that is, $Q=0$ or $U=0$), in complete contradiction to the null hypothesis that they should be uniformly distributed. It was argued that {\tt CLEAN} bias \citep{CLEAN} combined with small multiplicative and additive offsets were responsible for these effects, which can take values between $-90^{\circ}$ and $90^{\circ}$. Such small systematic effects are probably physically uninteresting in the context of astrophysical sources. However, the level of systematic control necessary in CMB observations is significantly higher and, therefore, it is interesting to consider whether this approach can be applied to the CMB. 
Application of similar techniques to CMB data presents some questions. Most importantly it is not clear whether the null hypothesis of a uniform distribution for $\alpha$ is valid. In Section~\ref{distribution} we will show that, although the mean of the histogram should the same for each bin, due to the inherent correlations in CMB maps the standard deviation about the mean will not be given by the usual $\sqrt{n_{i}}$ expected for a Poisson distribution, where $n_{i}$ is the number of pixels which have an $\alpha$ in the $i$-th bin. In principle this could be calculated analytically, but here we will just focus on estimating them numerically for specific cases. 

We will make the simplifying assumption that the observed Stokes' parameters $(Q_{\rm {obs}},U_{\rm {obs}})$ are just functions of the true values $(Q,U)$, that is, we will only allow for a restricted Muller matrix (see, for example, \citealt{systematic}). In this case we will see in Section~\ref{systematic} that it might be possible to detect systematic effects which correspond to global shearing of the polarization position angles. We believe that the basic method can be applied to more general situations, but will not be applicable to all possible systematic effects. However, we will also briefly consider the effect of a constant offset on Q and U.

The aim of this paper is to present the basic idea of the method. In addition we will apply it to the WMAP7 data \citep{seven_param}. We find that, even using foreground-reduced maps there is contamination in the Q and W bands, albeit at a level below the noise level in the maps. Further, we show that this contamination could be explained by a constant offset in the polarization. It should be possible to apply this technique to the upcoming Planck polarization data, which should have a significantly lower noise level \citepalias{Planck_blue} and have the possibility of detecting B-mode polarization \citep{Planck_B}.

\section{Statistics of CMB polarization angles}
\label{distribution}

In this section we will discuss the statistical properties of polarization position angles of the CMB. We will quote the basic results and rely on Monte-Carlo simulations to illustrate the point in specific cases.  

\subsection{Probability distribution of $\alpha$}
\label{p_alpha}

We will first consider the probability distribution of $\alpha$ in a single pixel before generalizing to a map with $N_{\rm {pix}}$ pixels (or a collection of $N_{\rm{pix}}$ individual polarization measurements). If the joint probability distribution for the Stokes' Q and U parameters, ${\cal P}(Q,U)$, is known, then one can convert from ${\cal P}(Q,U)$ to ${\cal P}(\alpha)$. Throughout this paper, we will be using the definitions $Q = P \cos 2\alpha$ and $U = P \sin 2\alpha$. In this analysis, we define Q and U by the standard definition relative to the pole of the map. In the simulations this is arbitrary, but when applied to WMAP data this is the galactic pole.

There are two steps required for the conversion. Firstly, we must calculate $ {\cal P}(P,\alpha)$. To do this, we use the standard formula for coordinate transformations which gives ${\cal P}(P,\alpha) = 2 P \mbox{ }{\cal P}[Q(P,\alpha),U(P,\alpha)]$. Then to  find ${\cal P}(\alpha)$, we need to integrate over all possible values of $P$, which gives
\begin{eqnarray}
\label{p_alpha_eqn}
{\cal P}(\alpha)=\int^{\infty}_{0}PdP\,{\cal P}\left[Q(P,\alpha),U(P,\alpha)\right] \, .
\end{eqnarray}

If Q and U are correlated, normally distributed random variables with zero mean, $\rm{Var}(Q)=\sigma_{QQ}$, $\rm{Var}(U)=\sigma_{UU}$, $\rm{Cov}(Q,U)=\sigma_{QU}$, where $\sigma_{QU}<\sqrt{\sigma_Q\sigma_U}$, then 
\begin{eqnarray}
{\cal P}(Q,U)&=&\frac{1}{2\pi\sqrt{\sigma_{QQ}\sigma_{UU}-\sigma_{QU}^2}} \\ 
\nonumber &\times& \exp\left(-\frac{Q^2\sigma_{UU}+U^2\sigma_{QQ}-2QU\sigma_{QU}}{2\left(\sigma_{QQ}\sigma_{UU}-\sigma_{QU}^2\right)}\right) \, .
\end{eqnarray}
We can substitute expressions for $Q$ and $U$ (in terms of $P$ and $\alpha$) in to this to obtain
\begin{eqnarray}
{\cal P}(Q(P,\alpha),U(P,\alpha)) &=& \frac{1}{4\pi\sqrt{\sigma_{QQ}\sigma_{UU}-\sigma_{QU}^2}} \\ 
\nonumber &\times& \exp\left(-P^2\frac{f_{\sigma}\left(\alpha\right)} {2\left(\sigma_{QQ}\sigma_{UU}-\sigma_{QU}^2\right)}\right) \, ,
\end{eqnarray}
where we have defined
\begin{eqnarray}
f_{\sigma} = \sigma_{QQ}+\sigma_{UU} + (\sigma_{UU}-\sigma_{QQ})\cos 4\alpha - 2\sigma_{QU}\sin 4\alpha \, ,
\end{eqnarray}
and, using eqn. \eqref{p_alpha_eqn}, we find that
\begin{eqnarray}
\label{p_alpha_noise_eqn}
{\cal P}(\alpha)=\frac{2\sqrt{\sigma_{QQ}\sigma_{UU}-\sigma_{QU}^2}} {\pi f_{\sigma}\left(\alpha\right)} \,.
\end{eqnarray}
In the case where $\sigma_{UU}=\sigma_{QQ}$ and $\sigma_{QU} = 0$ then ${\cal P}(\alpha)=1/\pi$, that is, the polarization position angle is uniformly distributed.

\subsection{Probability distribution in the multipixel case}

One can generalize this treatment to an ensemble of measurements. The measurements of $Q$ and $U$ at each of $N = N_{\rm pix}$ pixels will be treated as a set of random variables, $\mathbf{x}=(x_1,\ldots,x_{2N})=(Q_1,U_1,\ldots,Q_N,U_N)$. This set of variables have a multivariate Gaussian distribution
\begin{eqnarray}
{\cal P}(x_1,\ldots,x_{2N}) = \frac{e^{-\frac{1}{2}\left(\mathbf{x}-\mathbf{\mu}\right)^T\mathbf{\Sigma}^{-1} \left(\mathbf{x}-\mathbf{\mu}\right)}}{(2\pi)^{N}\left|\mathbf{\Sigma}\right|^{1/2}} \, ,
\end{eqnarray}
where $\mathbf{\Sigma}$ is the covariance matrix of $\mathbf{x}$ (which must not be singular). However, in this paper, we are dealing with the polarization position angle, and not Q and U directly. Thus, we need to calculate the distribution ${\cal P}(\alpha_1,\ldots,\alpha_N)$. In general this is non-trivial to calculate, since the values of $Q$ and $U$ can be correlated between pixels.

\subsubsection{Uncorrelated pixels}
\label{uncorrelated}

If we suppose that the pixels are not correlated, then the probability distribution is separable, so we have that ${\cal P}(\alpha_1,\ldots,\alpha_N) = {\cal P}(\alpha_1) \ldots {\cal P}(\alpha_N)$. This would be appropriate in the case of white noise where $C_{\ell} = \rm{const}$, which is equivalent to the value of each pixel being an independent, normally-distributed random variable. In this case, we can obtain the general solution by considering each pixel individually. Since we can only observe a finite number of pixels, we cannot observe the exact probability distribution, just some realisation of it. Instead, we form histograms by binning into different ranges for the position angle. However, note that what we are actually trying to measure is the underlying probability distribution, and a histogram is merely the simplest method for doing this when we have a finite number of observations. Using standard results, we find that, if the probability distribution for $\alpha$ is constant over all pixels, then
\begin{eqnarray}
\label{mean_N}
\langle n \rangle = N_{\rm{pix}} {\cal P}(\alpha_1 \leq \alpha < \alpha_2) \, ,
\end{eqnarray}
where $N_{\rm{pix}}$ is the total number of pixels and $n$ the number of pixels in the bin with $\alpha_1 \leq \alpha < \alpha_2$. We can also prove that, if the probability distribution for each pixel is different, we get that 
\begin{eqnarray}
\label{Ngen}
\langle n \rangle= \sum_{i=1}^{N_{\rm{pix}}} {\cal P}(\alpha_1 \leq \alpha_i < \alpha_2) \, .
\end{eqnarray} 
Furthermore, the standard deviation is given approximately by the formula $\sigma = \sqrt{n}$. It is also useful to define
\begin{eqnarray}
\label{n_hat}
\hat{n} = \frac{n- \bar{n}}{\bar{n}}\, ,
\end{eqnarray}
where we have defined $\bar{n} = N_{\rm{pix}} / N_{\rm{bins}}$, the average number of pixels in a bin. This number has the useful property that $\left \langle \hat{n} \right \rangle = 0$. The variance is approximately given by $\left \langle \hat{n}^2 \right \rangle = 1 / \bar{n}$.

\subsubsection{Correlated pixels}
\label{correlated}
 
For a more general $C_{\ell} \neq \rm{const}$ it is necessary to deal with correlations between pixels, and the results discussed in the previous section are modified. It can be shown that inter-pixel correlations do not affect the mean number of pixels in the bin and, thus, eqn. \eqref{Ngen} still gives the mean even for the correlated case. However, the probability distribution for $\alpha$ will be affected by correlations. In particular, this means that the simple $\sqrt{n}$ formula for the standard deviation does not in general apply to maps with correlated pixels. We find that the standard deviation of the number of points in each histogram bin depends on the particular CMB power spectrum used and, further, depends on the value of $\alpha$. To investigate the statistical distribution of $\alpha$ for general maps, we must therefore rely on simulations.

\begin{figure*}
\centering
\ifpdf \includegraphics[width=13cm,clip,viewport=20 250 600 805]{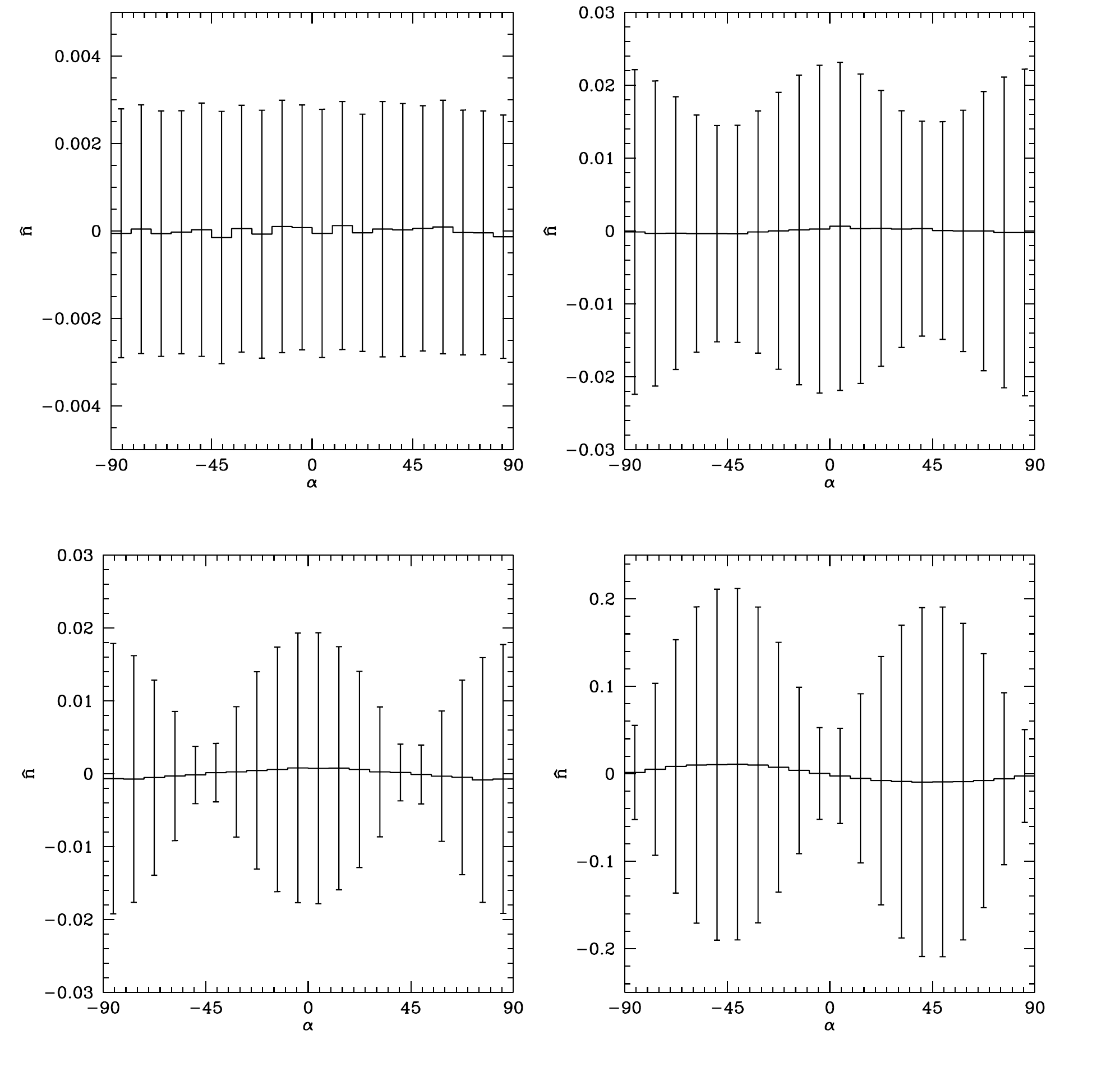}
\else  \includegraphics[width=14.5cm]{Diagrams/CMB_compare.eps} \fi
\caption{Histogram of the $\hat{n}$ in each bin for white noise ($C_{\ell} = \rm{const}.$) (top left) and three CMB-like power spectra, one with $r=1$ (top right), one with $C_{\ell}^{BB}=0$ (bottom left) and the final one with $C_{\ell}^{EE} = 0$ (and $r=1$) (bottom right). Note that the scales on the vertical axis are not the same. We find that, whilst the white noise case follows the standard $\sqrt{n}$ error formula, the CMB-like spectra have errors which are considerably larger, vary between bins and also depend on the particular power spectrum used as discussed in the text. The lower two plots appear to have an oscillation in the mean. However, this oscillation is within the expected standard deviation of the mean itself and, further, decreases with more realisations.}
\label{CMB_comp}
\end{figure*}

We have carried out simulations using four different power spectra, using a Monte-Carlo method with 1000 realisations with $N_{\rm side} = 512$ (which corresponds to $N_{\rm{pix}}~\approx~3~\times~10^6$). The four spectra are a constant-$C_{\ell}$ power spectrum (white noise) and three CMB spectra, one using a standard E-mode polarization spectrum and B-modes with $r=1$ (but no lensing), one using a standard E-mode spectrum with $r=0$ (no B-modes at all) and the final one using $r=1$ again, but with $C_{\ell}^{EE}$ set to 0 artificially to produce pure B-mode maps. The results are shown in Figure \ref{CMB_comp}. In all four cases used here, we find that $\langle \hat{n} \rangle$ is consistent with zero for all bins as discussed earlier.

From the histograms, however, it is obvious that the standard deviation is more complicated. The case of $C_{\ell}~=~\rm{const}$ follows the expected $\sqrt{n}$ law. However, the other three do not. The standard deviations in these are up to 10 times larger and, furthermore, they vary as a function of $\alpha$ and the $C_{\ell}$ values. We can describe these results qualitatively. From Figure \ref{CMB_comp}, we notice that a power spectrum that is dominated by E-modes (such as the CMB itself) has errors that oscillate according to
\begin{eqnarray}
\sigma(\alpha) = A_1 + B_1 \cos 4\alpha \, ,
\end{eqnarray}
where $A_1$ and $B_1 > 0$ are measurable parameters. Similarly, it appears that power spectra dominated by B-modes oscillate following 
\begin{eqnarray}
\sigma(\alpha) = A_2 - B_2 \cos 4\alpha \, ,
\end{eqnarray}
With $B_2 > 0$. In the case where both are equal, the oscillating term disappears, and the error becomes constant. Overall, it appears that we can write 
\begin{eqnarray}
\sigma(\alpha) = A + B \cos 4\alpha \, ,
\end{eqnarray}
where $A$ and $B$ are both functions of the particular power spectrum used.

From these results, we can see that the standard deviation for a pure B-mode CMB spectrum is considerably larger (factor of $\sim 10$) than that for either the pure-E or the combined E and B spectra. The combined E and B spectrum has a similar maximum standard deviation to the pure-E spectrum. However, the minimum standard deviation is considerably larger due to the addition of the B-modes. A CMB spectrum with $r \ll 1$ will have a similar error distribution to the pure E-mode case (since B-modes are then negligible).

\begin{figure}
\ifpdf \includegraphics[height=6.5cm,clip, viewport=0 0 560 550]{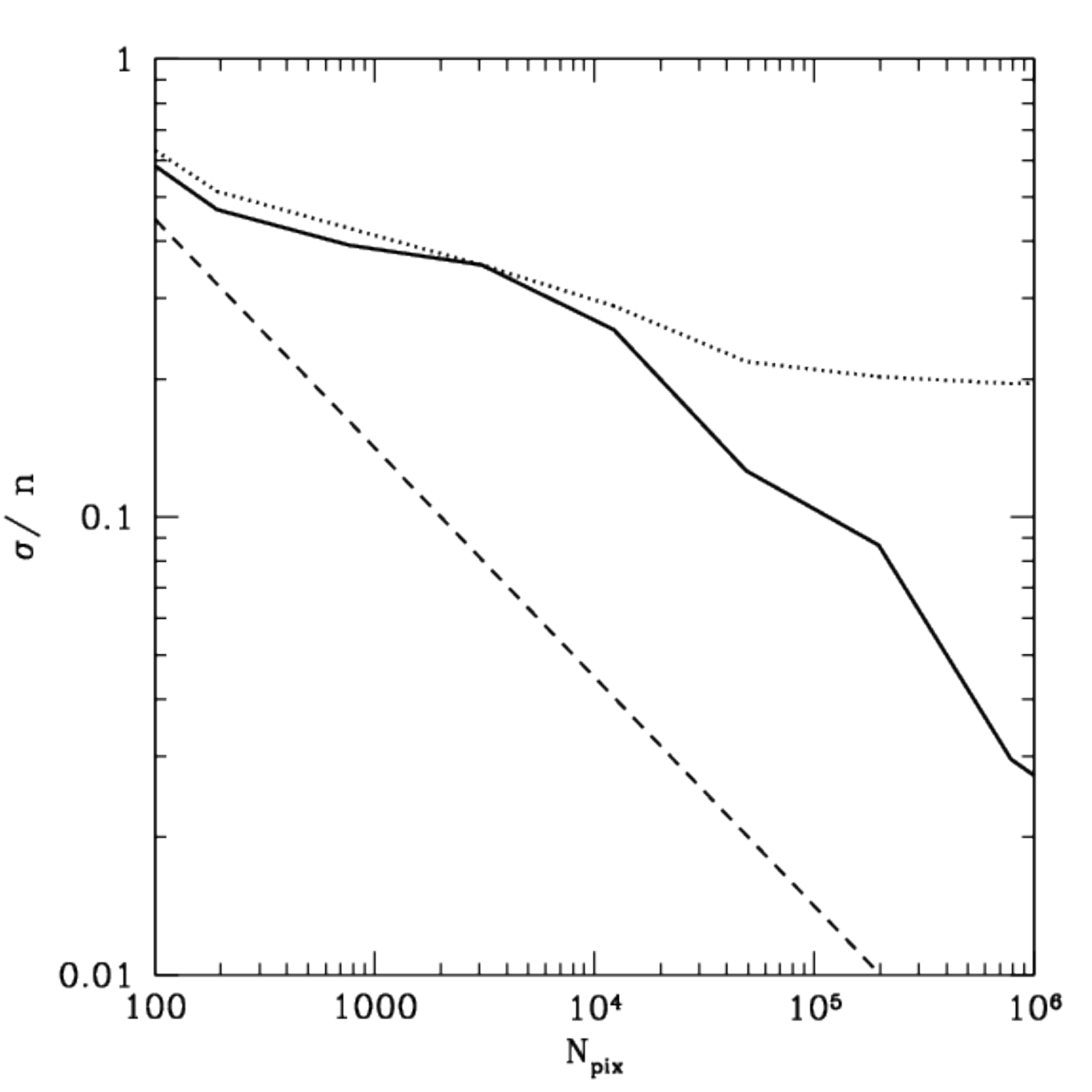}
\else \includegraphics[width=16.5cm]{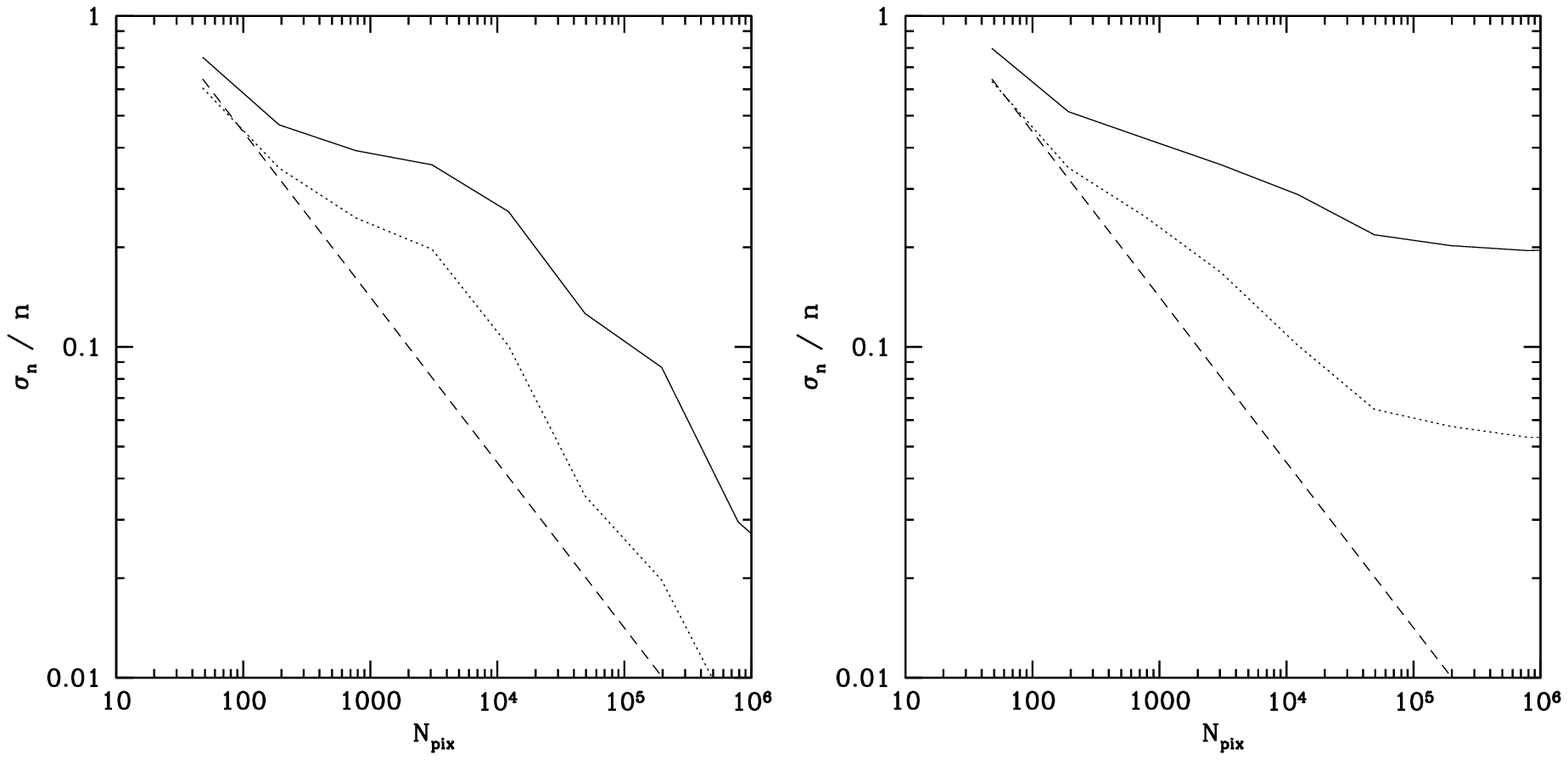} \fi			
\caption{Fractional error, $\sigma / n$, in the number of pixels in the bin with the highest standard deviation against $N_{\rm{pix}}$ (used here as a proxy for the resolution) for pure E-mode and pure B-mode CMB runs. The dashed line is $\frac{1}{\bar{n}}$, the fractional error for uncorrelated pixels, the whole line shows a pure E-mode run and the dotted line shows a pure B-mode run.}
\label{npix}
\end{figure}

The increase in the standard deviation of the histogram bins as compared to the uncorrelated case is due to the effect of the inter-pixel correlations. These correlations act to reduce the effective number of pixels being measured and, thus, increase the level of variance. We have plotted the maximum standard deviation for different values of $N_{\rm{pix}}$ for a pure-E mode CMB and for a pure-B mode CMB in Figure \ref{npix}. These results show a substantial difference between the two cases. Whereas the standard deviation on the pure-E modes changes approximately as $1/\sqrt{N_{{\rm pix}}}$, the standard deviation for the B-mode case hardly falls at all after $\sim 4 \times 10^4$ pixels. This is because the E-mode polarization has more power on small scales than the primordial B-mode polarization (excluding lensing), which is dominated by large scales. As a result, increasing the resolution beyond a certain point in the latter case just results in creating many highly-correlated clusters of pixels, meaning that the effective number of pixels (and, thus, the standard deviation in the histogram bins) asymptotes. However, were the effects of lensing on the B-mode spectrum included, this effect would become less prominent, due to the small-scale nature of the induced fluctuations.

\section{Effect of systematic errors}
\label{systematic}

The previous section gives some idea of the properties of histograms generated from considering the polarization position angle of both CMB signals and noise. However, the main purpose of this paper is to consider how well systematic errors can be detected using position angle histograms, and so we will now consider the effects of various types of systematic error on $\alpha$. 

\subsection{Scaling, rotation and shear systematics}
\label{shear}
\label{properties}

The first type of systematic we will consider corresponds to scalings, rotations and shears of the Stokes parameters Q and U, represented by
\begin{eqnarray}
\label{simple_system}
\left(\begin{array}{c}
Q_{\rm obs} \\
U_{\rm obs}
\end{array} \right) =
\left(\begin{array}{c}
Q \\
U
\end{array} \right) +
\left(\begin{array}{cc}
\epsilon_{QQ} & \epsilon_{QU} \\
\epsilon_{UQ} & \epsilon_{UU}
\end{array} \right)
\left(\begin{array}{c}
Q \\
U
\end{array} \right) \, ,
\end{eqnarray} 
where it is expected that $\epsilon_{ij}$ will be small and are assumed to be constant over the map.

A more natural way to represent this form of systematic is to write
\begin{eqnarray}
\label{system_param}
\left(\begin{array}{cc}
\epsilon_{QQ} & \epsilon_{QU} \\
\epsilon_{UQ} & \epsilon_{UU}
\end{array} \right) 
&=& \epsilon_0 \left(\begin{array}{cc} \cos \theta & -\sin \theta \\ \sin \theta & \cos \theta \end{array}\right) \\
\nonumber &+& \epsilon_1  \left(\begin{array}{cc} 1 & 0 \\ 0 & -1 \end{array}\right)
+ \epsilon_2  \left(\begin{array}{cc} 0 & 1 \\ 1 & 0 \end{array}\right) \, ,
\end{eqnarray}
where
\begin{eqnarray}
\tan \theta &=& \frac{\epsilon_{UQ} - \epsilon_{QU}}{\epsilon_{QQ} + \epsilon_{UU}} \, , \\
\epsilon_0 &=& \frac{1}{2} \sqrt{\left(\epsilon_{QQ}+\epsilon_{UU}\right)^2 + \left(\epsilon_{UQ}-\epsilon_{QU}\right)^2} \, , \\
\epsilon_1 &=& \epsilon_{QQ} - \epsilon_{UU} \, , \\
\epsilon_2 &=& \epsilon_{QU} + \epsilon_{UQ} \, .
\end{eqnarray}

These parameters all have a physical meaning: $\epsilon_0$ corresponds to an overall scaling; $\theta$ to a global rotation of the position angles;  $\epsilon_1$ and $\epsilon_2$ to shearing. If $\epsilon_0$ and $\theta$ are non-zero there is no effect on the probability distribution for $\alpha$: $\epsilon_0 \neq 0$ does not change $\alpha$ and $\theta \neq 0$ just translates all values of $\alpha$. However, $\epsilon_1$ and $\epsilon_2$ have an effect which it might be possible to detect.

\begin{figure*}
\ifpdf \includegraphics[width=13cm,clip,viewport=25 500 600 755]{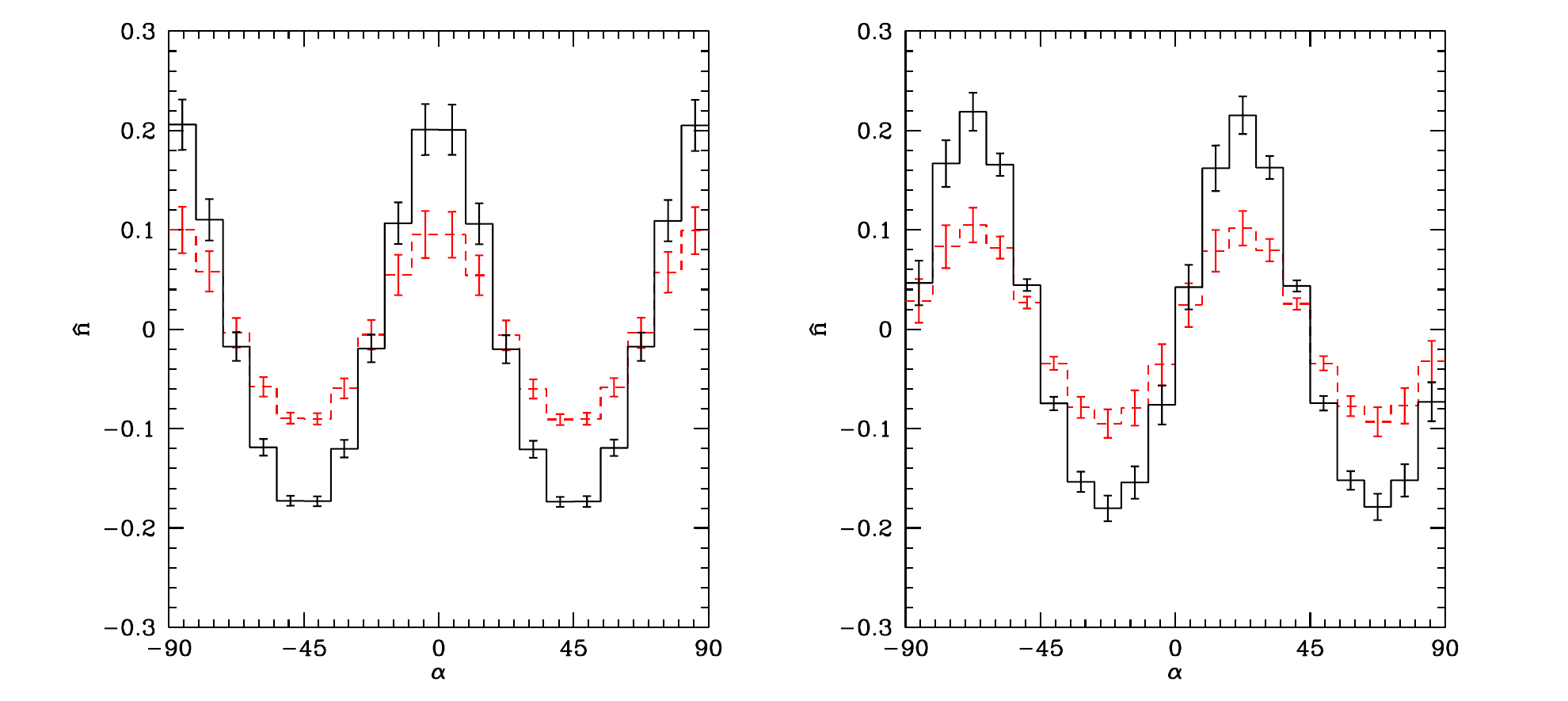}
\else  \includegraphics[width=14.5cm]{Diagrams/CMB_systematic.eps} \fi
\caption{Histogram of $\hat{n}$ against $\alpha$ including systematic effects added to CMB realisations with $r=0.1$ for different values of $\epsilon_1$ (left) and $\epsilon_2$ (right). In each histogram, the solid black line shows $\epsilon_{1,2} = 0.1$, and the dashed red line shows $\epsilon_{1,2} = 0.05$. We estimate that it should be possible to detect values of $\epsilon_1$ or $\epsilon_2$ of around $0.01$.}
\label{epsilon}
\end{figure*}

In order to investigate the impact of $\epsilon_1$ and $\epsilon_2$, we have applied different systematic errors to CMB maps, setting one of $\epsilon_1$ or $\epsilon_2$ to be non-zero in each case. We have then plotted histograms of the results in Figure \ref{epsilon}. In each case we have created 100 CMB maps with $r~=~0.1$

In the presence of such systematic effects, a flat initial probability distribution for $\alpha$ is, to first order, transformed into the following
\begin{eqnarray}
\label{pe}
{\cal P}(\alpha) = \frac{1}{\pi}(1 + 2\epsilon_1 \cos 4\alpha + 2\epsilon_2 \sin 4\alpha) \, ,
\end{eqnarray}

\subsection{Additive systematics}
\label{additive}

Another possible systematic effect that we could imagine is one whereby the values of $Q$ and $U$ are shifted by some constant value, such that
\begin{eqnarray}
\label{const_def_Q}
Q_{\rm obs} = Q + \delta Q \,, \\
\label{const_def_U}
U_{\rm obs} = U + \delta U \,.
\end{eqnarray}

\begin{figure*}
\ifpdf \includegraphics[width=13cm,clip,viewport=25 500 600 755]{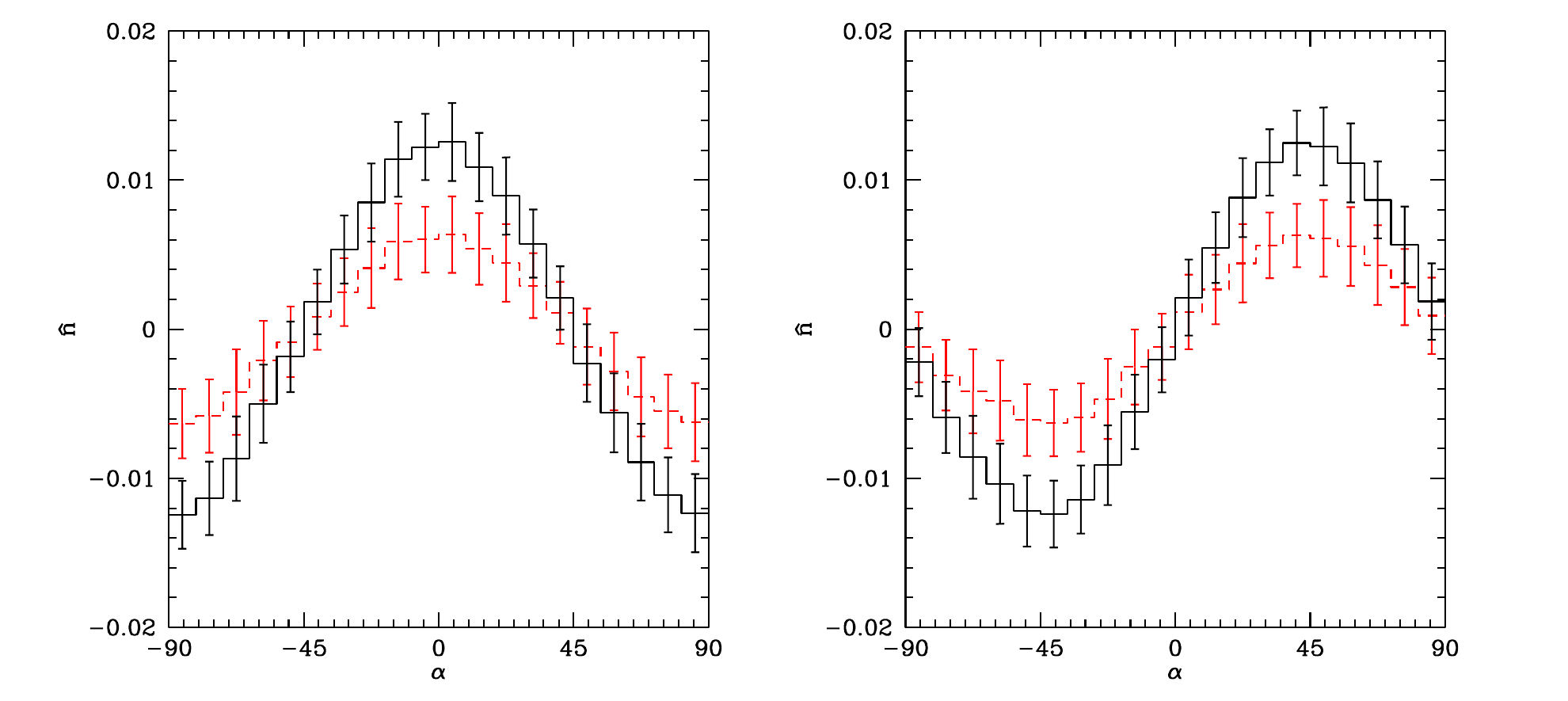}
\else  \includegraphics[width=14.5cm]{Diagrams/systematic_add.eps} \fi
\caption{Histogram of $\hat{n}$ against $\alpha$ including systematic effects for different values of $\delta Q$ (left) and $\delta U$ (right) added to a white-noise signal, using a resolution of $N_{\rm side}=512$. In each histogram, the solid black line shows a systematic with $\sqrt{\delta Q^2 + \delta U^2} / P = 0.01$, and the dashed red line shows a systematic with $\sqrt{\delta Q^2 + \delta U^2} / P = 5 \times 10^{-3}$ of the input signal. We estimate that it should be possible to detect this type of systematic error down to a level of around $2 \times 10^{-3}$ of the polarized noise.}
\label{additive_fig}
\end{figure*}

In order to understand the effect of these errors, we have performed several simulations, adding various levels of systematic to a white noise signal. Some sample results are shown in Figure \ref{additive_fig}. It can clearly be seen that the result of applying such systematic errors is that $\hat{n}$ seems to have an oscillation frequency  of $2 \alpha$, rather than $4 \alpha$ as with the case considered in Section \ref{shear}, with the probability distribution of $\alpha$ given by the formula
\begin{eqnarray}
\label{p_add}
{\cal P}(\alpha) = \frac{1}{\pi}\left( 1 + \frac{\delta Q \cos 2\alpha + \delta U \sin 2\alpha}{\left\langle P \right\rangle}\right) \, .
\end{eqnarray}

This pattern could therefore be a clear indication of systematics of this form. Note, however, that this only applies where the systematic added is constant over all pixels. A random additive systematic may result in a different pattern, as might a signal caused by leakage from the intensity channel.

\subsection{Detection of systematic errors in experiments}
\label{NC_chi_square}

Using the properties of the $\chi^2$ distribution, it is possible to determine the likelihood of detecting a systematic at a given confidence level. It can be shown that, to have a 50\% chance of a detection at a 95\% confidence level, we need $\epsilon_1~=~5.5 \times 10^{-3}$ and $\epsilon_2~=~7.8 \times 10^{-3}$, and for a 90\% chance of detection, we require $\epsilon_1~=~8.0 \times 10^{-3}$ and $\epsilon_2~=~1.1 \times 10^{-2}$. Notice that systematic errors with $\epsilon_1~\neq~0$ are slightly easier to detect than those with $\epsilon_2~\neq~0$. This is because the oscillations in $\epsilon_1$ are in phase with those in the errors, which means that the minimum error coincides with the point at which the oscillations in ${\cal P}(\alpha)$ are largest, whereas in the case of $\epsilon_2~\neq~0$, the minimum error occurs where there is no difference between ${\cal P}(\alpha)$ in the control and test cases.

We have also carried out a similar analysis for additive systematics. In this case, we find that a 50\% chance of a detection at a 95\% confidence level will require $\sqrt{\delta Q^2 + \delta U^2} / P = 2.0 \times 10^{-3}$, and a 90\% chance will require $\sqrt{\delta Q^2 + \delta U^2} / P = 2.9 \times 10^{-3}$. Note that the level of systematic required for detection is lower than for the shear systematics shown above because of the use of white noise as the base signal rather than CMB.

\section{Test on WMAP 7}
\label{WMAP}

\subsection{Properties of WMAP polarization measurements}
\label{WMAP_properties}

In this section, we will test our ideas on the WMAP seven-year polarization maps (see \citet{seven_prelim} for further details). The full ($N_{\rm side} = 512$) WMAP polarization maps for each difference assembly (DA) are strongly dominated by noise and foregrounds, with the CMB signal being barely detectable. It is possible to obtain a map which contains a larger proportion of CMB signal by combining the measurements for each DA at a given frequency, but the resultant maps are still noise and foreground dominated at the single pixel level. It is also possible to reduce the amount of foreground contamination by producing foreground-reduced maps (see \citet{seven_prelim} for details) and by masking the galactic plane, where most of the foreground contamination is found. The WMAP experiment has 5 frequency bands - K band (23 GHz), Ka band (33 GHz), Q band (41 GHz), V band (61 GHz) and W band (94 GHz), but the foreground-reduced maps only exist for Ka, Q, V and W bands, since the K band is used to construct templates for the foreground removal.

In order to conduct a test, there are two steps. First we simulate the WMAP noise maps and determine the expected distribution (and the error in this distribution). Then we compare the predicted distribution with the actual WMAP data, allowing us to assess the level of any discrepancy. The first of these steps will be discussed in this section, and the second will be discussed in Section \ref{WMAP_data}. 

\label{WMAP_simulate}

In order to detect any systematic errors in the WMAP data, we must first consider what we would expect to see in the absence of systematics. Since the maps are noise-dominated, eqn. \eqref{p_alpha_noise_eqn} should give the probability distribution for $\alpha$ for any given pixel, once the noise correlation matrix for that pixel is known. However, unlike in Section \ref{uncorrelated}, the noise is not constant across the sky, since each pixel has been measured a different number of times \citep{third_observe}. 

An $N_{\rm{obs}}$ matrix is provided with the WMAP data set, and this, together with the given noise level for each DA, allows the noise correlation matrix for each pixel to be found (see \citet{WMAP_supp} and \citet{third_observe} for more details). These can then be combined together to produce noise correlation matrices for the combined frequency map. Once this is known, eqn. \eqref{Ngen} could in theory be used to determine the mean number of pixels in each histogram bin, by summing the probabilities of each pixel being in the bin. In practice, it is simpler to carry out Monte Carlo simulations in order to determine the mean and variance of the number of pixels in each bin.

We have generated 1000 noise realisations with the correct statistical properties, added to a CMB realisation with $r=0.1$ (our results are not sensitive to the value) to each realisation and then calculated a histogram of $\hat{n}$ in each bin. 

\subsection{Comparison of simulations to real data}
\label{WMAP_data}

Once we have simulations of the WMAP noise, we then need to compare the results to those generated from the actual WMAP maps. WMAP has a total of 10 difference assemblies at five different frequencies, and each DA has nine years of data currently available. However, for the purpose of this paper we will consider only the seven-year data, both with and without foreground reduction applied, and with and without a mask applied.

For each frequency band and with each type of map, we have produced combined maps using the process outlined in \citet{third_observe}. We then generated histograms by plotting, for each bin,
\begin{equation}
\hat{n}_{\rm{sim}} = \frac{n - n_{\rm{sim}}}{\bar{n}} \, ,
\end{equation}
where $n_{\rm sim}$ is the average number of pixels in the bin from simulations, and $\bar{n} = N_{\rm{pix}} / N_{\rm{bins}}$. 

\begin{figure*}
\centering
\ifpdf \includegraphics[width=17cm,clip,viewport=20 270 570 830]{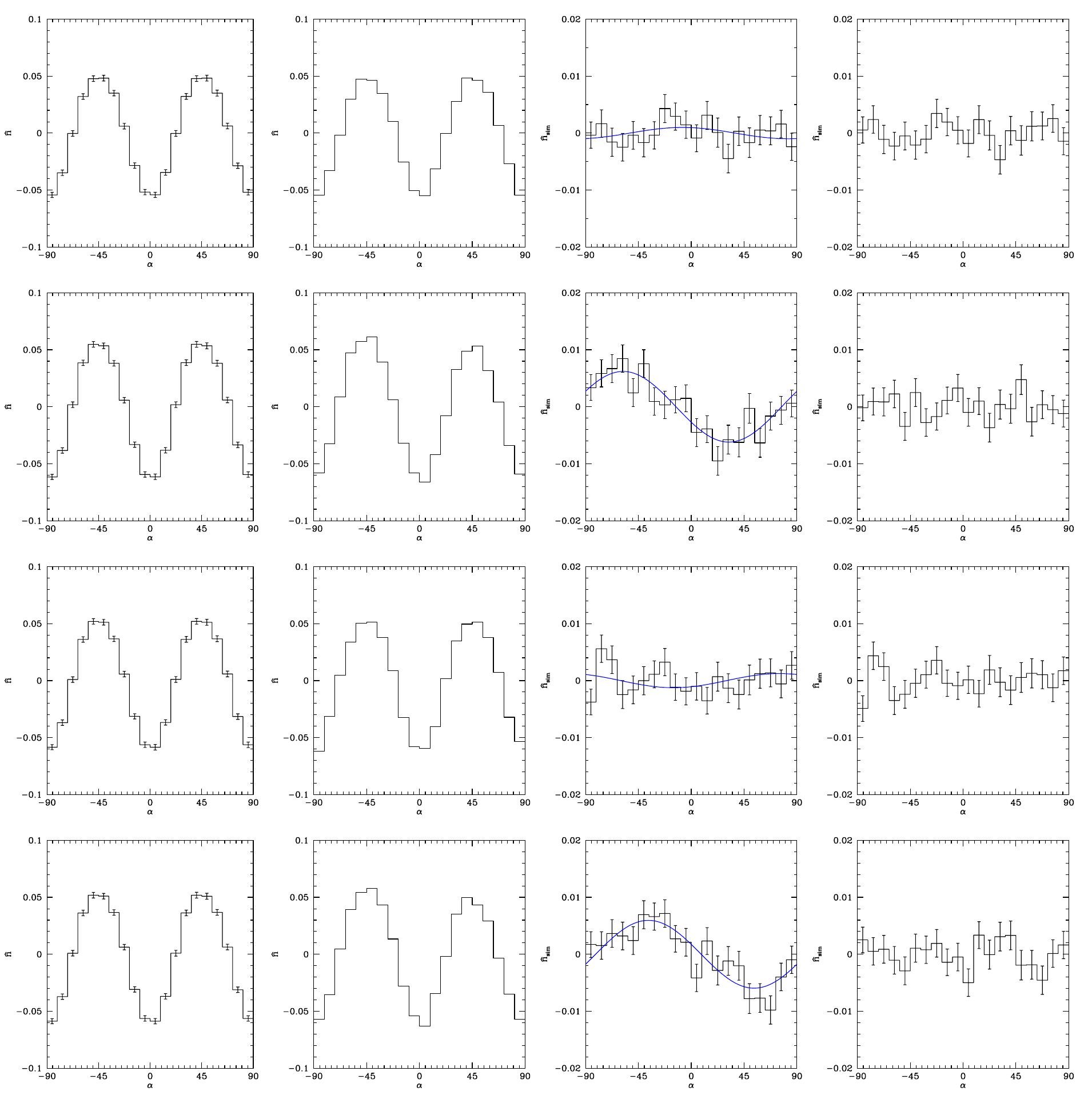}
\else  \includegraphics[width=15.5cm]{Diagrams/WMAP/WMAP_compare_redu_unmask.eps} \fi
\caption{Histograms of $\hat{n}$ and $\hat{n}_{\rm sim}$ vs. $\alpha$ for the four highest frequency WMAP bands, Ka, Q, V and W (from top to bottom) using foreground-reduced maps with no masking. The four histograms shown in each row are, from left to right, simulations based on the WMAP7 noise maps, the actual WMAP data, the difference between the two and the difference minus the best-fit sinusoidal model. There is a clear discrepancy between the observed and simulated data in the Q and W bands, which appears to oscillate as a function of $2 \alpha$.}
\label{WMAP_redu_unmask}
\end{figure*}

\begin{figure*}
\centering
\ifpdf \includegraphics[width=17cm,clip,viewport=20 270 570 830]{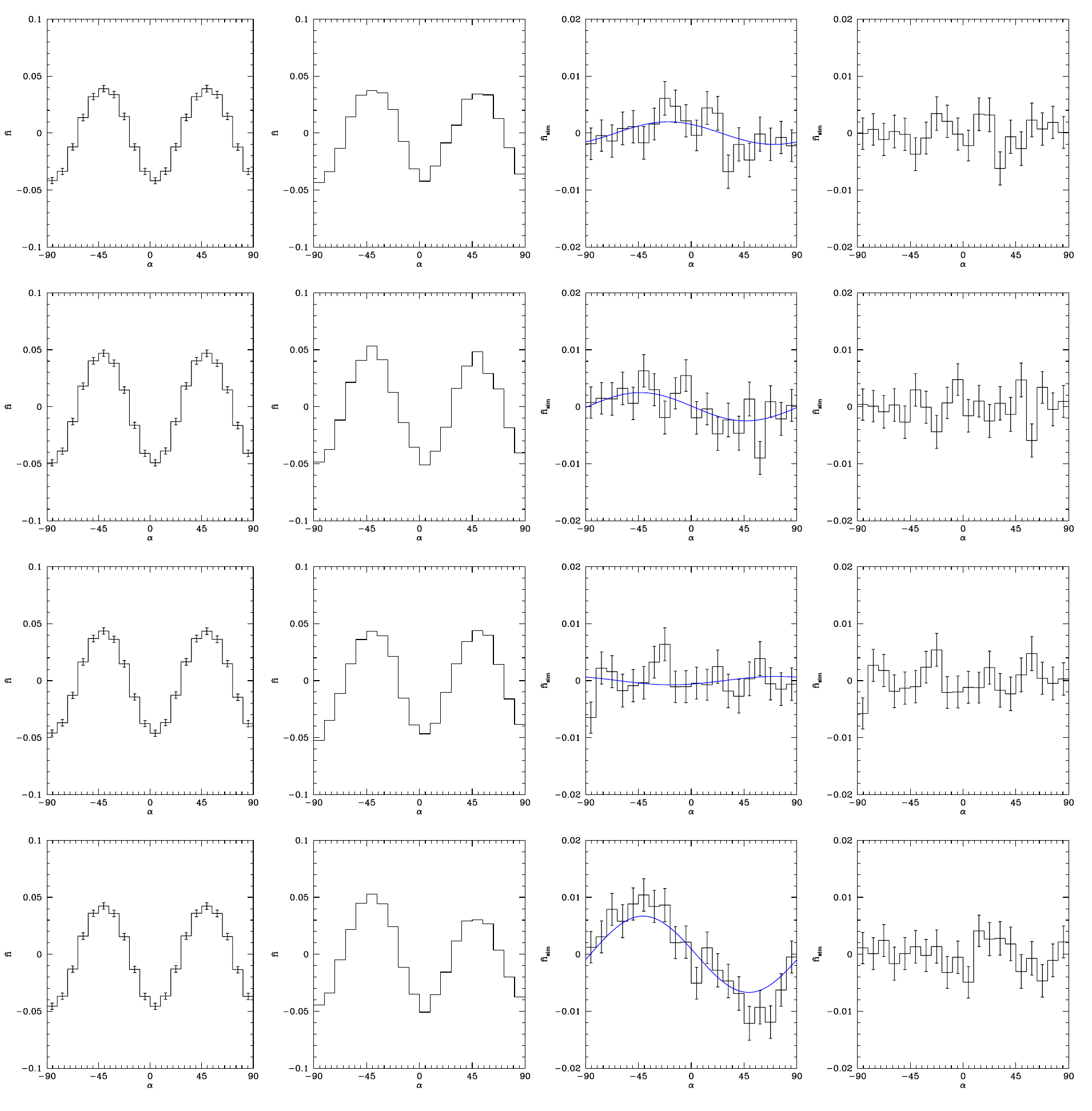}
\else  \includegraphics[width=15.5cm]{Diagrams/WMAP/WMAP_compare_redu_mask.eps} \fi
\caption{Histograms of $\hat{n}$ and $\hat{n}_{\rm sim}$ vs. $\alpha$ for the four highest frequency WMAP bands, Ka, Q, V and W (from top to bottom) using foreground-reduced maps with masking also applied. The four histograms shown in each row are, from left to right, simulations based on the WMAP7 noise maps, the actual WMAP data, the difference between the two and the difference minus the best-fit sinusoidal model. Whilst the osscilation in the Q band data is less clear than in the masked case, it appears to have a similar phase, suggesting it is still a genuine signal.}
\label{WMAP_redu_masked}
\end{figure*}

Without any attempt to mitigate for galactic foregrounds, the histograms are dominated by the effects of the North Polar Spur where $\alpha \approx 0$. Therefore, we have concentrated our analysis on the foreground reduced WMAP maps. The resultant histograms are shown in Figures \ref{WMAP_redu_unmask} and \ref{WMAP_redu_masked}.

There is a clear indication of a sinusoidal oscillation with frequency $2 \alpha$ in the W band data, and also in the unmasked Q band data. However, in order to determine for certain if these results are consistent with chance, we need to look at the $\chi^2$ values, which are presented in Table \ref{chi_tab}. For the W band, and for the unmasked Q band, there is conclusive evidence that the divergence of the histogram from zero is not purely random. In addition, there is strong but not conclusive evidence that there is some divergence in the masked Q band data. It is possible to fit the results to a function of the form $A \sin [2(\alpha + \theta)]$, and the best-fit sinusoid, as determined using the method given below, is plotted for comparison.
\small
\begin{table*}
\small
\begin{tabular}{ccccc}
\hline
\hline
Band & Unmasked reduced $\chi^2$ value & ${\cal P}(\chi^2 > \chi^2_{\rm obs})$ & Masked reduced $\chi^2$ value & ${\cal P}(\chi^2 > \chi^2_{\rm{obs}})$ \\
\hline
Ka & $0.750$ & $0.770$ & $1.188$ & $0.257$ \\ 
Q & $4.080$ & $3.9 \times 10^{-9}$ & $1.558$ & $0.057$ \\ 
V & $1.071$ & $0.374$ & $0.931$ & $0.543$ \\
W & $3.942$ & $1.3 \times 10^{-8}$ & $6.128$ & $5 \times 10^{-15}$ \\
\hline
\hline
\end{tabular}
\caption{Table of reduced $\chi^2$ values (using 19 degrees of freedom) for the difference between the seven-year WMAP foreground reduced maps and simulations of the noise and CMB signal expected in these maps. It is clear that the null hypothesis is not valid for the W band data or the unmasked Q band.}
\label{chi_tab}
\label{WMAP_chi_square}
\end{table*}
\normalsize

\label{sin}

The best fit sinusoid, and the $\chi^2$ value associated with it, are shown in Tables \ref{chi_fit_unmask} (for the unmasked case) and \ref{chi_fit_mask} (for the masked case). In all cases, fitting a sinusoid to the histogram results in a $\chi^2$ which is consistent with chance. However, the V band and the unmasked Ka band do not show any significant reduction in the $\chi^2$ relative to the case without fitting (as shown in Table \ref{chi_tab}) so there is no evidence of a significant effect in these maps. For the masked Ka band, whilst there is a substantial improvement, the $\chi^2$ without fitting is still entirely consistent with chance, so we cannot reject the hypothesis that no fit is necessary in this case. For the remaining frequencies, however, there is clear evidence of a oscillation of the form described above, with the parameters given in the table.

Additionally, in the case of the W band, the best-fit amplitude and $\theta$ in the masked and unmasked case are equal to within the uncertainties. This suggests that the cause of this error is some effect that is common to these two cases. 


\small
\begin{table*}
\small
\begin{tabular}{ccccc}
\hline
\hline
Band & Best fit $\theta$ & Best fit amplitude& Reduced $\chi^2$ value & ${\cal P}(\chi^2 > \chi^2_{obs})$ \\
\hline
Ka & $53 \pm 23^{\circ}$ & $\left(1.0 \pm 0.8\right) \times 10^{-3}$ & $0.719$ & $0.795$ \\ 
Q & $-77 \pm 4 ^{\circ}$ & $\left(6.2 \pm 0.8\right) \times 10^{-3}$ & $0.853$ & $0.638$ \\ 
V & $-60 \pm 18 ^{\circ}$ & $\left(1.2 \pm 0.8\right) \times 10^{-3}$ & $0.995$ & $0.462$ \\
W & $-82 \pm 4^{\circ}$ & $\left(6.0 \pm 0.8\right) \times 10^{-3}$ & $1.043$ & $0.406$ \\
\hline
\hline
\end{tabular}
\caption{Table of reduced $\chi^2$ values (using 19 degrees of freedom) for the difference between the unmasked seven-year WMAP foreground reduced maps and simulations of the noise and CMB signal expected in these maps fitted to a sinusoid. We see that the Q and W bands have a discrepancy of around $6 \times 10^{-3}$ of the polarized intensity in the map.}
\label{chi_fit_unmask}
\label{WMAP_chi_fit_unmask}
\end{table*}

\begin{table*}
\small
\begin{tabular}{ccccc}
\hline
\hline
Band & Best fit $\theta$ & Best fit amplitude& Reduced $\chi^2$ value & ${\cal P}(\chi^2 > \chi^2_{obs})$ \\
\hline
Ka & $65 \pm 13^{\circ}$ & $\left(2.0 \pm 0.9\right) \times 10^{-3}$ & $0.752$ & $0.759$ \\ 
Q & $89 \pm 10^{\circ}$ & $\left(2.5 \pm 0.9\right) \times 10^{-3}$ & $0.953$ & $0.513$ \\ 
V & $62 \pm 35^{\circ}$ & $\left(7 \pm 9\right) \times 10^{-4}$ & $0.912$ & $0.563$ \\
W & $86 \pm 4^{\circ}$ & $\left(6.7 \pm 0.9\right) \times 10^{-3}$& $0.828$ & $0.669$ \\
\hline
\hline
\end{tabular}
\caption{Table of reduced $\chi^2$ values (using 19 degrees of freedom) for the difference between the masked seven-year WMAP foreground reduced maps and simulations of the noise and CMB signal expected in these maps fitted to a sinusoid. There is a hint of a discrepancy in the Q band, although it is weaker than in the unmasked case, and a clear discrepancy in the W band.}
\label{chi_fit_mask}
\label{WMAP_chi_fit_mask}
\end{table*}
\normalsize

\section{Discussion and Conclusions}
\label{discuss}

In this paper, we have put forward a novel test of CMB polarization maps using histograms of the polarization position angle. We have shown that, for the case where the pixels are uncorrelated, such as for a noise-dominated map, the standard $\sqrt{n}$ rule will apply to the uncertainty in the number of pixels in the bin. However, for the CMB, which has inter-pixel correlations, this will be higher. We have not analytically characterised the properties of CMB maps, but we can see that, as would be expected from symmetry, the probability distribution of the position angles for a CMB map is flat, and this can be analytically proven. Additionally, we have characterised the properties of histograms of CMB maps. We find that the errors oscillate as a function of $\cos 4 \alpha$, with the exact co-efficients depending on the nature of the power spectrum. We also find that the error is substantially larger for B-mode-only maps, likely due to the fact that much of the power in the B-mode map used is at large scales, and thus pixels are highly correlated. 

We have then considered various forms of systematic error, and their detection. Our primary focus was on systematic errors that are functions of the true values of $Q$ and $U$. We found that, using this method, systematic errors which have the effect of rotating or stretching the Stokes parameters could not be detected. However, shear-like systematic errors could be detected at the $0.5\%$ level in a CMB-dominated map, and show a distinct sinusoidal oscillation with a period of $4 \alpha$. This calculation does, however, assume no inter-pixel correlations, which is manifestly not true for the CMB. Thus, further investiation of this effect will be necessary. Additionally, we did a brief investigation of additive systematics. These were shown to also be detectable, to below the $0.5\%$ level. However, unlike the previously-considered systematics, these have oscillations with a period of $2 \alpha$.

Finally, as a test of the method, we applied it to the WMAP 7 year data. Here, we found clear evidence of residual foreground contamination even after masking was applied to the non-foreground-reduced data, although this is not unexpected. The more interesting results, however, are in the foreground-reduced data. The Ka and V band data appears to be clean, but for the W and Q bands there is evidence of an effect, which can be fitted with a sinusoid with a period of $2 \alpha$. For the W band, the effect is consistent between the masked and unmasked data, but for the Q band it is somewhat weaker in the masked data. This may suggest that some of the effect can be attributed to foreground contamination, but not all of it, however there is no concrete evidence of this. Alternately, it is possible that the foreground removal process used to produce the foreground reduced maps is the cause of the observed signal.

\label{foregrounds}

\begin{figure*}
\ifpdf \includegraphics[width=17cm,clip,viewport=0 530 600 805]{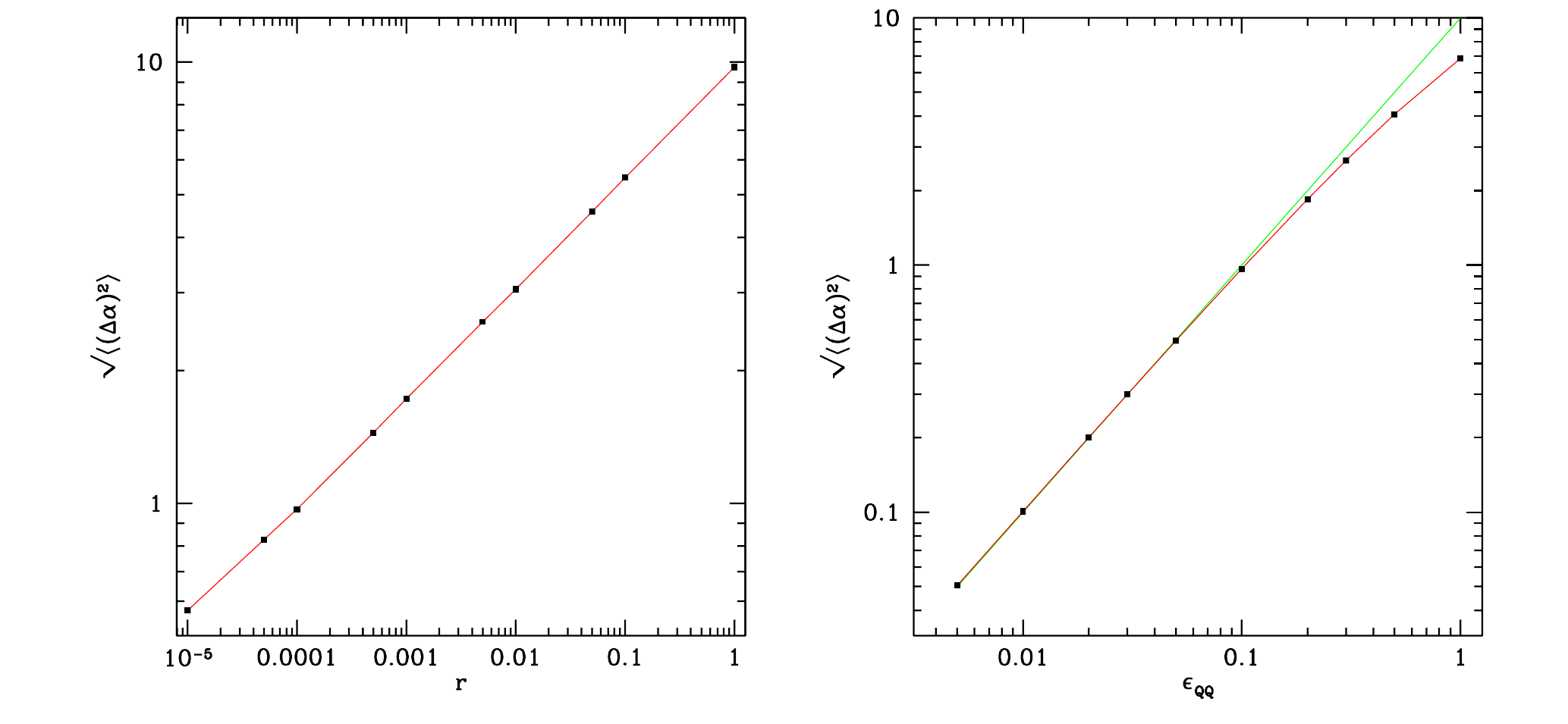}
\else  \includegraphics[width=16.5cm]{Diagrams/RMS(alpha).eps} \fi
\caption{r.m.s. difference in $\alpha$ vs. $r$ (left) and $\epsilon_{QQ}$ (right). The red lines are the simulated values, and the green lines (not visible on the left hand plot due to absolute agreement with the simulations) are the formulae given in eqn. \eqref{RMS_alpha}.}
\label{RMS(alpha)}
\end{figure*}

In order to consider the usefulness of our method, we need to know how these systematic errors affect B-mode measurements. An indication of the effect of systemetics on B-mode measurements can be obtained by comparing the r.m.s. change in $\alpha$ in the case of the addition of B-modes to the change caused by the addition of systematics. The results are shown in Figure \ref{RMS(alpha)}. Firstly, we have looked at the change between a pure-E mode CMB realisation, and the same realisation but with B-modes added, using simulations with $N_{\rm side} = 512$. We found that there is a simple formula connecting $\sqrt{\left \langle \left(\Delta\alpha\right)^2 \right \rangle}$ with $r$, where $\Delta \alpha$ is the change in $\alpha$ between the pure E-mode realisation and the realisation with E and B-modes. It is also possible to compare a CMB realisation both with and without systematic errors, and we have found an approximate formula for this situation. Although we have only considered $\epsilon_{QQ}$ here, these results will apply equally to the other terms in the Muller matrix. Plots for these two cases are shown in Figure \ref{RMS(alpha)}.

Based on these results, we can then find empirical relations giving $\sqrt{\left \langle \Delta \alpha^2 \right \rangle}$ as a function of $r$ and $\epsilon_{QQ}$
\begin{eqnarray}
\label{RMS_alpha}
\epsilon_{QQ} \approx r^{\frac{1}{4}} \approx \frac{1}{10} \left \langle \Delta \alpha^2 \right \rangle^{\frac{1}{2}} \, .
\end{eqnarray}
From this, it can be seen that a value of $\epsilon_{QQ} = 0.1$ causes an equivalent change in the r.m.s. of $\alpha$ to the addition of B-modes with $r = 10^{-4}$. This is indicative of the level at which this systematic would affect B-mode measurements.

\section{Acknowledgements}

We would like to thank Michael Brown for helpful comments on the manuscript, and Clive Dickinson and Patrick Leahy for valuable discussions on the topic of the paper.









\bibliographystyle{mn2e_fix}
\bibliography{paper}
%
%
%
%
%

\end{document}